Review Article

7858 words

# Visibility retrieval in Michelson wide-field stellar interferometry

I. MONTILLA*, J. SELLOS, S. F. PEREIRA, and J. J. M. BRAAT

Delft University of Technology, Imaging Science and Technology Department, Optics Research Group, P.O. Box 5046, NL-2600 GA Delft, The Netherlands

*Corresponding author. Email: iciar.montilla@obs.univ-lyon1.fr

Wide field interferometry has become a subject of increasing interest in the recent years. New methods have been suggested in order to avoid the drawbacks of the standard wide-field method (homothetic mapping) which is not applicable when the aperture is highly diluted; for this reason imaging with non-homothetic arrays is being extensively studied [1], [2]. The field of view of a pupil plane interferometer or a densified array consists only of a few resolution elements; in order to improve these systems, we developed a new method consisting of a Michelson pupil-plane combination scheme where a wide field of view can be achieved in one shot. This technique, called "staircase mirror" approach, has been described in a previous paper [3] and uses a stair-shaped mirror in the intermediate image plane of each telescope in the array, allowing for simultaneous correction of the differential delay for both the on- and off-axis image positions. Experimental results have been obtained showing the simultaneous recovering of the fringes of off-axis stars with an appreciable angular separation, and with a contrast similar to that of the on-axis reference star. With this example, we demonstrate an increase of the field of view by a factor of five, with no need of extra observation time. In this article, we present a further analysis of the method. We investigate how to retrieve the visibility when a star is focused on the edge of a step of the stair-shaped mirror. Even though the Optical Pathlength Difference (*OPD*) correction is discontinuous, we show both numerically and analytically that the visibility can be completely recovered, so that no information is lost. Our experimental results demonstrate that the visibility can be retrieved to within a 1% error.

# 1 Introduction

In the field of optical interferometry several aspects need to be improved or newly implemented. One of these aspects is the extended field of view: what can be done in order to get a wide interferometric field of view and what are the possible applications of such a goal.

A wide field of view is very important in applications such as the observation of extended or multiple objects or the fringe acquisition and/or tracking on a nearby unresolved object [4], and also for the not-trivial task of reducing the observation time. For ground-based arrays the field of view should at least be equal to the isoplanatic patch [5].

Many observational studies in astronomy such as studies on galaxy formation and kinematics, star formation, stellar evolution and circumstellar physics as well as astrometric detection of extra-solar planets and binaries require a large field of view.

As explained in a previous paper [3], to measure at least the contrast of the fringes, the *OPD* error over the entire array field has to be smaller than the coherence length. In an interferometer, this correction is done for the pointing direction of the telescope, but for any other angle the *OPD* will be different. If this difference is larger than the coherence length, coherent interference of the beams will not take place. Therefore, the interferometric field of view is limited by the spectral resolution, $\lambda_0/\Delta\lambda$, with $\lambda_0$ being the central wavelength and $\Delta\lambda$ the bandwidth. In fact, it is the product of the spectral and the spatial resolution, $\lambda_0/B$ [6]. For example, the field of view of an interferometer with a baseline of 200 meters observing at 2.2 μm with a spectral resolution of 10 is approximately 23 milliarcsec. A low spectral resolution also produces an effect called "bandwidth smearing" [7]. The interferometer observes a finite bandwidth $\Delta\lambda$, but the external geometrical *OPD* is compensated for the central wavelength $\lambda_0$. The averaging of the visibility over the bandwidth produces a radial blurring of the image, the so-called "bandwidth smearing", a sort of chromatic aberration. One way to increase the field of view is to increase the spectral resolution, but a high spectral resolution may be incompatible with the observation of faint objects. It is thus desirable to keep a moderate spectral resolution, and to find alternative ways to increase the field of view.

There are two main technological concepts to observe a large field of view, depending on the type of beam combination: homothetic mapping for multiaxial beam combiners, and wide field mosaic imaging for coaxial beam combiners. In homothetic mapping, the configuration of the telescopes as seen from the science source is re-imaged to a smaller scale, but maintaining orientation and relative separations, before the beams interfere. As sketched in Fig. 1, a Fizeau-type instrument is intrinsically a homothetic mapper, where the beam combination scheme has a natural wide field- of-view only limited by atmospheric anisoplanatism and the correcting adaptive optics system. A Michelson-type telescope array can also be used as a homothetic mapper when the images are recorded in the focal plane and if the exit pupil after the telescopes is an exact demagnified replica of the input pupil as seen by the incoming wavefront. Pupil rotation has to be accurately controlled to maintain the orientation of the exit pupils. For example, to cophase the longest baselines of the Very Large Telescope Interferometer (VLTI) to better than 300 nm over a continuous field of view of 4 arcsec the baseline has to be known with an accuracy of a few tens of microns and pupil rotation to within 8 arcsec [8]. The interferometer then behaves like one huge telescope of which only the fraction of its surface that contains the telescopes is being used. This is the case of the Large Binocular Telescope Interferometer, which actually has a continuous field of view of 1 arcmin operating at 2.2 μm. Homothetic mapping has been anticipated by the designers of the VLTI Laboratory by providing a pit with a diameter of 2 meters in the interferometric laboratory. This technique is also being developed in the Delft Testbed Interferometer (DTI) [9] The DTI is a Fizeau-type interferometer designed to acquire an imaging angle proportional to 2 arcsec within the VLTI set-up.

Wide field mosaic imaging is being developed at the Wide-field Imaging Interferometry Testbed [10]. The technique is analogous to the mosaicing method employed in radio astronomy [11], adapted to a Michelson pupil-plane beam combiner with detection in the image plane. A delay line is used to scan the optical path length through the sky and a NxN pixels array detector records simultaneously the temporal fringe patterns from many adjacent telescope fields. The recorded data needs to be jointly deconvolved to reconstruct the image. If a detector with a large number of pixels is used, and the image plane is sampled at high enough spatial frequency, then this technique could in principle be used to multiply the field size by a factor of N/2, reaching a field of several arcmin. This technique is under development at NASA for the Space Infrared Interferometric Telescope (SPIRIT) and the Submillimeter Probe of the Evolution of Cosmic Structures (SPECS).

The drawbacks of the above mentioned techniques are that on one side, homothetic mapping requires a highly complex positioning system and very accurate calibration of the optical parameters of the instrument, and on the other side, wide field mosaicing requires long observation time since wide field of view is not acquired in one shot.

In order to avoid these drawbacks we developed a new approach to the pupil recombination problem, i.e., a system that uses a Michelson pupil-plane scheme with a wide field of view to be acquired in one shot [3],[12]. We introduced a new technique, called "staircase mirror" approach, which consists of positioning a stair-shaped mirror in an intermediate image plane for each telescope in the array. In this way, the light from stars arriving at different angles on the telescopes is imaged on different steps on the staircase mirror (see Fig. 2). Consequently, the extra path length introduced by the staircase mirror provides in a first approximation simultaneous correction of the *OPD*. The size and height of the steps on the mirror depends on the baseline and the pointing direction of the telescopes. During the observation, as the entrance pupil geometry varies with the astronomer's hour angle, the height of the steps must be adapted in order to follow the change in optical path length. Given the discontinuous nature of the staircase mirror, it may appear that the correction of the *OPD* in such a way would yield a discontinuous field of view. When an object is focused on the edge of a step, the light is reflected from two steps simultaneously, implying a different phase shift for each part. The effect on the detected light distribution after combining the beams is that instead of having one fringe set, with one maximum at the zero path position, there are two fringe sets, each with a relative maximum. Each set has a relative contrast, and the addition of the contrast of the two components of the fringe system results in the expected contrast of the source multiplied by a factor that depends on the bandwidth and the baseline; with the aid of this factor the visibility can be retrieved.

In this paper we focus on the retrieval of the visibility in the Michelson case with discontinuous path length compensation in order to compare the analytical results with the experimental data and validate the concept. This paper is organised as follows: the general analytical description of the effect of the staircase mirror on the visibility is assessed in Section 2, beginning with the general case (Section 2.1) followed by the specific case where the source is focussed on the edge of the step of the staircase mirror (Section 2.2). The latter analysis is essential in order to demonstrate that a continuous field of view can be obtained with our method. Further, the algorithm to retrieve the visibility of a source that is focused in the edge of a step is presented in Section 2.3. A description of the experimental set-up built to verify our wide-field concept is given in Section 3.1, and in Section 3.2 and 3.3 experimental results are shown demonstrating that by using the staircase mirror and the algorithm presented in this paper the visibility component of the celestial object can be completely recovered. Although not necessary for the experiment that we show here, we also have performed a theoretical analysis of the effect due to the diffraction of the edge of the mirror that may occur if the dimension of the recollimating optical system after the staircase mirror is comparable with the diffracted beam from the staircase mirror. This analysis is presented in Appendix A.

**2 Analytical description**

*2.1 General case*

The system we use for our theoretical analysis is represented in Fig. 3 and consists of a two-arm Michelson interferometer. At the entrance of each arm there is an aperture with a lens. In the focal plane of the lens in one of the arms we place the staircase device formed by steps of depth *d*. The beams from each arms are collimated and recombined. Further we suppose a source that is observed with one of the arms of the interferometer, with an aperture of radius $\rho_A$, numerical aperture $N_A$ and transmission function *T* constant over the spectral bandwidth, i.e., a "top-hat" bandwidth of $\Delta\lambda$ centered on $\lambda_0$. The flux [13], in $Wm^{-2}$, at the focal plane, with focal length $F$, of the objective is:

$$D_A(x, y) = \Delta\lambda \left(\frac{\pi N_A \rho_A}{\lambda_0}\right)^2 T^2 P(x, y), \quad (1)$$

where $P(x, y)$ depends on the source geometry (a point source, a binary, a disk) and is equal to the convolution of the flux of the object with the normalized Point Spread Function (*PSF*) of the system. In our narrow-band approximation ($\Delta\lambda \ll \lambda_0$), we suppose that $P(x,y)$ is given by its shape at the central wavelength $\lambda_0$. A similar function is obtained for the flux at the focal plane of the other arm. In order to make the analysis simpler, we consider the same aperture radius and numerical aperture for the combined focus. If the entrance apertures of the arms of the interferometer are separated by a distance B, assuming their aperture radii, numerical apertures and transmission functions to be equal, the flux $D(x, y, t)$ detected in the combined focal plane is given by [14]

$$D(x, y, t) = 2D_a(x, y)\left[1 + \mu(B)\cos(\varphi(x, y, t))\text{sinc}\left(\frac{\pi(\varphi(x, y, t))}{\kappa_0 L_c}\right)\right], \quad (2)$$

where $\mu(B)$ is the modulus of the complex coherence factor (visibility) of the source observed with baseline B, $L_c = \lambda_0^2/\Delta\lambda$ is the coherence length and $\kappa_0 = 2\pi/\lambda_0$ is the wave number at the central wavelength $\lambda_0$. The phase difference between both arms has two contributions: the external geometrical optical path difference and the time-dependent phase introduced by the temporal modulation of the internal optical path:

$$\varphi(x, y, t) = \varphi_{ext}(x, y) + \varphi_t \qquad (3)$$

The external phase difference is given by the coordinates of the baseline vector $\mathbf{B} = (x_B, y_B)$ and the coordinates defining the position of the point in the focal plane

$$\varphi_{ext}(x, y) = \frac{\kappa(x_B x + y_B y)}{F} \qquad (4)$$

When observing with non-monochromatic light with coherence length $L_c$, and scanning a maximum path that is a factor $a$ of the coherence length $\Delta t_{max} = aL_c$, coherent interference does not occur for the points with $\kappa\varphi_{ext}(x, y) > aL_c$ and the visibility can not be retrieved.

Our aim is to correct this differential *OPD*;. this is done by introducing an extra internal Optical Path Difference at the focal plane. For this purpose, a staircase-shaped function is chosen. The internal phase function introduced by a stair-shaped mirror with (2N+1) steps of width $w$ and depth $d$ placed in the focal plane is given by

$$\varphi_{int}(x, y) = \sum_{n_s = -N}^{N} \kappa n_s d \, \text{rect}\left(\frac{x_B x + y_B y}{w\mathrm{B}} - n_s\right) \qquad (5)$$

with the function $\text{rect}(x)$ given by $\text{rect}(x) = 1$ when $|x| \leq 1/2$ and zero elsewhere.

With the staircase mirror in the focal plane of arm *A'*, the phase difference between both arms is:

$$\varphi(x, y, t) = \varphi_{ext}(x, y) - \varphi_{int}(x, y) + \varphi_t \qquad (6)$$

The external *OPD* is completely corrected for the source points focused in the middle of the steps, where $x_B x + y_B y = n_s w$ with $n_s = 0, \pm 1, \pm 2, ..., \pm N$. For the points not located in the middle of the steps, the external *OPD* is not completely corrected, but the difference is always smaller than the maximum path scanned $\Delta t_{max}$, assuring partially coherent interference for the entire field determined by the field of view and the number of steps of the staircase mirror.

The power $L(t)$, in units of Watts (W), is obtained by simply integrating the flux. The time dependence in $L(t)$ is due to the temporal modulation of the optical path. The visibility $\mu(B)$ is calculated using the relation

$$\mu(B) = \frac{L_{max} - L_{min}}{L_{max} + L_{min}}, \qquad (7)$$

where $L_{max}$ and $L_{min}$ are the maximum and minimum powers, obtained when the temporal modulation of the internal optical path has a value such that the phase $\varphi_t = 0$ or $\pm\pi$, respectively.

The relative phase, $\varphi$, that gives the position of the off-axis star with respect to the on-axis one, can be calculated by measuring the distance from the peak of the off-axis fringe pattern to the peak of the on-axis pattern, $n_{p\text{-}p}$, and knowing in which step the off-axis star is focused:

$$\varphi = \frac{1}{B}\left(2nd\cos\alpha + n_{p-p}\right), \qquad (8)$$

where $\alpha$ is the angle that the mirror is forming with the optical axis, and $n_{p\text{-}p}$ can be both positive and negative.

### *2.2 Source focused on the edge of a step*

We now describe the fringe power corresponding to a source focused on the edge of a step, as shown in Fig. 4. When the focal image is exactly centred on the edge of the step a wavelength-dependent phase shift of $\delta(\lambda)$ is imparted to half the Fourier spectrum of the corresponding Point

Spread Function (*PSF*). In this case, the interference of the beams results in two sets of fringes, both with a contrast smaller than the one corresponding to the star focused in the middle of the step, as can be seen in the simulations plotted in Fig. 5. In this situation, the internal phase introduced by the steps is not the same across the entire *PSF*. Given a step depth of *d* and considering that the normal to the surface of the mirror forms and angle $\alpha$ with the optical axis, we have that the phase difference introduced by the step is given by $\kappa_0 d_{eff}$, where $d_{eff} = 2d \cos \alpha$. The light reflected from each step will result in two relative maxima of the power: the first one occurs when the phase is 0 and $-\kappa_0 d_{eff}$ for the light focused on the first and second step, respectively, and the second one when the phase is $\kappa_0 d_{eff}$ and 0 for the light focused on the first and second step, respectively. Correspondingly, there will also be two relative minima. The resulting fringe pattern comprises two sets of fringes, each of them with visibilities $\mu_1$ (B) and $\mu_2$ (B), respectively. We will first calculate $\mu_1$ (B); the maximum of this fringe set is reached when $\varphi_t = 0$ in the first step and $\varphi_t = -\kappa_0 d_{eff}$ in the second one. Considering that the steps are orthogonal to the *x*-axis and that the edge of the step is located at $x_2$, the integration of the flux gives a maximum power equal to:

$$L_{max} = G \left\{ \left[ 1 + \mu(\mathrm{B}) \right] \int_{x_1}^{x_2} \int_{y_1}^{y_2} P(x,y) dx dy \right.$$
$$\left. + \left[ 1 + \mu(\mathrm{B}) \cos(\kappa_0 d_{eff}) E \right] \int_{x_2}^{x_3} \int_{y_1}^{y_2} P(x,y) dx dy \right\}, \quad (9)$$

where:

$$G = 2\Delta\lambda \left( \frac{\pi N_a \rho_a}{\lambda_0} \right)^2 T^2; \quad E = \mathrm{sinc}\left( \frac{\pi d_{eff}}{L_c} \right). \quad (10)$$

For practical reasons, the integration domain $(x_1, x_3)$ and $(y_1, y_2)$ should be limited to the size of the diffraction image of the single aperture. In reality, the effective integration area is determined by the detector, so a sufficiently small photosensitive area has to be chosen depending on the size of the *PSF* of the system. The minimum of the fringe set occurs when $\varphi_t = \pm\pi$ in the first step and $\varphi_t = \pm\pi - \kappa_0 d_{eff}$ in the second one. The minimum power is:

$$L_{\min\pm} = G\left\{\left[1-\mu(\mathrm{B})A\right]\int_{x_1}^{x_2}\int_{y_1}^{y_2}P(x,y)dxdy \right.$$
$$\left. +\left[1-\mu(\mathrm{B})\cos(\kappa_0 d_{eff})C_{\pm}\right]\int_{x_2}^{x_3}\int_{y_1}^{y_2}P(x,y)dxdy\right\}, \quad (11)$$

with:

$$A = \mathrm{sinc}\left(\frac{\pi\lambda_0}{2L_c}\right); \quad C_{\pm} = \mathrm{sinc}\left(\frac{\pi d_{eff}}{L_c} \mp \frac{\pi\lambda_0}{2L_c}\right). \quad (12)$$

The relative visibility $\mu_1(\mathrm{B})$ calculated from Eq. (7) equals:

$$\mu_1(\mathrm{B}) = \mu(\mathrm{B})\left\{\frac{(1+A)\int_{x_1}^{x_2}\int_{y_1}^{y_2}P(x,y)dxdy + \cos(\kappa_0 d_{eff})(E+C_{\pm})\int_{x_2}^{x_3}\int_{y_1}^{y_2}P(x,y)dxdy}{2\int_{x_1}^{x_3}\int_{y_1}^{y_2}P(x,y)dxdy + \mu(\mathrm{B})\left[(1-A)\int_{x_1}^{x_2}\int_{y_1}^{y_2}P(x,y)dxdy + \cos(\kappa_0 d_{eff})(E-C_{\pm})\int_{x_2}^{x_3}\int_{y_1}^{y_2}P(x,y)dxdy\right]}\right\}. \quad (13)$$

Similarly we calculate the visibility of the second fringe set $\mu_2(\mathrm{B})$, where the maximum is reached when $\varphi_t = \kappa_0 d_{eff}$ in the first step and $\varphi_t = 0$ in the second one:

$$\mu_2(B) = \mu(B) \left\{ \frac{\cos(\kappa_0 d_{eff})(E+C_\pm)\int_{x_1}^{x_2}\int_{y_1}^{y_2} P(x,y)dxdy + (1+A)\int_{x_2}^{x_3}\int_{y_1}^{y_2} P(x,y)dxdy}{2\int_{x_1}^{x_3}\int_{y_1}^{y_2} P(x,y)dxdy + \mu(B)\left[\cos(\kappa_0 d_{eff})(E-C_\pm)\int_{x_1}^{x_2}\int_{y_1}^{y_2} P(x,y)dxdy + (1-A)\int_{x_2}^{x_3}\int_{y_1}^{y_2} P(x,y)dxdy\right]} \right\}. \quad (14)$$

For typical values used in aperture synthesis, we have, in general, $(1-A) \ll 1$ and $\cos(\kappa_0 d_{eff})(E-C_\pm) \ll 1$. These approximations are validated, for example, with the parameters of the tabletop set-up described in Section 3; for a bandwidth $\Delta\lambda = 150$ nm and a central wavelength $\lambda_0 = 575$ nm, we have that $(1-A) \cong 0.02$. The value of $\cos(\kappa_0 d_{eff})(E-C_\pm)$ for the same spectral range, for a baseline of 30 mm is approximately 0.0025. In this way, we can neglect the second term in the denominator of Eqs. (13) and (14). Adding $\mu_1(B)$ and $\mu_2(B)$, we obtain:

$$\mu_1(B) + \mu_2(B) \cong f\mu(B), \quad (15)$$

where $f$, the so-called edge factor, is given by

$$f = \frac{f_+ + f_-}{2}; \quad f_\pm = \frac{1+A+\cos(\kappa_0 d_{eff})(E+C_\pm)}{2}. \quad (16)$$

Combining the measured fringe visibilities $\mu_1(B)$ and $\mu_2(B)$, and calculating the edge factor using the experimental parameters, the expected visibility can be recovered. As can be seen from Eq. (10) and (12), the edge factor depends on the central wavelength, the coherence length and the depth of the steps. As the depth of the steps depends on the projected baseline, for a given central wavelength and a given coherence length, the edge factor will fluctuate during observation as the baseline projection changes. In Fig. 6 the edge factor is plotted as a function of the baseline, for a maximum baseline of 35 mm. Note that the maximum change in the visibility is approximately 15% of the expected value.

The relative angle, $\varphi$, that gives the position of the star focused on the edge with respect to a reference one, can be calculated by measuring the distance from one of the peaks of the fringe sets to the peak of the reference fringe pattern. If $n_{p_1-p_0}$ is the distance from the peak of the first fringe set to the reference fringe pattern, and $n_{p_2-p_0}$ the distance from the peak of the second fringe set to the reference fringe pattern, and if we know between which steps, $n$ and $n+1$, the star is focused, the relative angle is

$$\varphi = \frac{1}{B}\left(nd_{eff} + n_{p_1-p_0}\right) = \frac{1}{B}\left((n+1)d_{eff} - n_{p_2-p_0}\right), \tag{17}$$

where $n_{p_1-p_0}$ and $n_{p_2-p_0}$ can be either positive or negative but always have opposite sign.

In this section we have not considered the effect that the diffraction produced by the edge of the step has on the calculated visibility, but a study in detail can be found in Appendix A. Using Fourier analysis we have determined the limits of the approximation presented in this section, showing that it is valid when the re-collimating lens is larger in lateral size than 1.5-1.7 times the output pupil, in which case the error in the calculated visibility is smaller than 2%. For our experiments the re-collimating lens was approximately 15 times larger than the output pupil, therefore diffraction effects can be neglected

## 3 Description of the experiment and results

### *3.1 Description of the experiment*

In order to demonstrate the ideas presented in Section 2, we designed and implemented a tabletop set-up consisting of a two-telescope Michelson-type interferometer. In the focal plane of one of the arms of the interferometer a staircase mirror was placed. A scheme of the set-up is plotted in Fig. 7 and it consists of three main blocks: the star simulator, the interferometer and the beam combiner.

The star simulator is a Xenon arc lamp with a set of filters that reduces its spectrum to a bandwidth of 150 nm (500-650 nm), resulting in a coherence length of 2.2 μm. The light is imaged onto the star mask, which consists of a set of pinholes of 5 μm, where each pinhole defines one star. In the experiment described here the illuminated mask has four pinholes, distributed in a square with separation of 150 μm. The size of the holes ensures full spatial coherence for the beams over the entire spectral bandwidth. The light from two pinholes is not coherent, ensuring two independent sources as is the case with real stars. The light from the star mask is collimated in such a way that at the entrance pupil of the telescopes there are four beams, simulating four distant objects in the sky. Two of these objects have an external *OPD* equal to zero, therefore they are considered on-axis objects. The other two have an external *OPD* different from zero, hence for the interferometer they are off-axis objects with an off-axis angle of 1.03 arcmin.

The two interferometer beams are obtained by wavefront division with the aid of two apertures of 20 mm diameter, and the separation between the apertures ("baseline") is 30 mm. After each aperture a telescope objective is placed, and at the focal plane of one of the interferometer arms, a staircase mirror (see Fig.2), is set. The staircase mirror introduces a temporal delay to the imaged spots that depends on the incidence angle of the incoming beam at the entrance pupil of the telescope and on the tilt angle of the mirror. We have chosen a tilt of $45^{o}$ so that the symmetry between both arms is easily maintained. The width of the step depends on the effective focal length, *F*, of the telescope objective. The parameters were chosen with the criterium that the point spread function of the focused spot was several times smaller than the width of the step. In the actual configuration the effective focal length is 6.563 m, and the system is free of lateral and axial chromatic aberration. The depth of the step *d* is related to the width *w* by:

$$\frac{d}{w} = \frac{B}{2F}, \tag{18}$$

where B is the baseline or separation between the telescopes. The staircase mirror was manufactured at Philips Research Laboratories (Eindhoven, The Netherlands) with a diamond turning machine (DTM) using a chisel with radius of curvature of 3 μm. For the chosen step width of 1339 μm both the on- and the off- axis beams are focused close to the centre of their respective steps. The measured value of the depth of the steps is 2.65 μm.

The staircase mirror holder can be rotated, so that the steps can be set either perpendicular to the baseline, in order to correct the external optical path, or parallel to it, which corresponds to the situation when no staircase mirror would be present, i.e., no simultaneous correction of the *OPD* for the on- and off-axis beams.

The light is collimated again before entering the delay line and beam combiner block. Two mirrors on top of a piezo stage, set in one of the interferometer arms, provide a delay line to modulate the optical path. The beams overlap at a beam splitter/combiner cube. The two outputs of the beam splitter are detected by a low-noise photometric detector and a CCD camera that measures the modulated fringes and controls the pupil position, respectively.

The piezo stage of the delay line was actuated with a triangle function of 4 volts amplitude, equivalent to a modulation of the path of approximately 9 μm. The fringes produced by the coherent interference of the beams were detected with the photodetector, as well as the power of each beam for calibration of the visibility. In this way, the fringes corresponding to the four stars in the star mask were obtained.

*3.2 Experimental results: general case*

The data shown in this section concerns the case when the focussed beams on the staircase mirror corresponding to the on-and off-axis stars does not fall on the edge of the steps, so that one fringe envelope per star is obtained. Also, in order to compare the effect of the staircase mirror on the visibility with the case when the staircase mirror would be a flat mirror, we performed two sets of measurements. In the first case, we consider no external *OPD* correction for off-axis stars. This can be achieved by keeping all elements of the experiment the same except that the staircase mirror is rotated in a way that the steps are aligned parallel to the baseline. The corresponding fringes in this case are shown in Fig.8, with a total scanning distance of 9 µm. The fringe from the on-axis star (Fig. 8a) has a visibility of 0.82±0.02. The experimental limitation on the value of the visibility is attributed to dispersion effects of the achromats and by the fact that the star mask is not a perfect point source. The fringe from the off-axis star (Fig. 8b) has a much lower visibility of 0.14±0.02 since it is not in the field of view of the interferometer. Similar results are found with the other two on- and off-axis stars. In the second case, we place the staircase mirror with the steps perpendicular to the baseline, so that correction of the external *OPD* does take place. The results are shown in Fig. 9. The visibility calculated from the calibrated fringe of the two on-axis stars is again 0.82±0.02 (Fig. 9a and 9c). The visibility of the two off-axis stars is 0.74±0.02 (Fig. 9b and 9d). This value, as compared with the visibility of the off-axis star without correction (Fig. 8b), clearly demonstrates the effect of the *OPD* correction due to the staircase mirror.

The relative angle, $\varphi$, that gives the position of the off-axis star with respect to the on-axis one, can be calculated by measuring the distance from the peak of the off-axis fringe pattern to the peak of the on-axis pattern, $n_{p-p}$, and using Eq.(8). Here we obtain the value of $(3.3\pm0.1)\times10^{-4}$ rad for the relative angular separation between the stars, which closely agrees with the calculation of the phase by direct measurement of the experimental parameters.

The visibility of the off-axis fringes is not equal to the visibility of the on-axis fringe pattern again attributed to the dispersion effects.[15]

### *3.3 Experimental results: spot on the edge*

In this section, we demonstrate the effect on the fringe visibility when a spot falls on the edge of the one step on the staircase mirror. Experimentally, this is done with the same set-up as in Section 3.2 except that the mask is slightly displaced in a way that the light from the star is focussed near or between two steps on the staircase mirror. The fringes were measured for five different positions of a star with respect to the edge of the step. First, the star was situated on-axis and at the center of one step, obtaining one set of reference fringes. The visibility of this set of fringes was used as reference. Then, the star mask was translated in such a way that the star was placed close enough to the edge so that a second fringe set started to appear. Subsequently, the star was moved to the edge between the steps, resulting in two sets of fringes with similar contrast. As the star is moved further towards the center of the next step, the contrast of one of the sets increases while the other one decreases in a way that the total contrast stays constant. When the star is focused close to the center of the next step, the interference fringes form one set again. These experiments were performed for two different baselines and are shown below.

### 3.3.1 Experimental results with a 30 mm baseline

The measured calibrated fringes are shown in Figs. 10 and 11 for two different central wavelengths and bandwidths. Because the edge factor depends on the bandwidth and central wavelength of the detected light, measurements were taken with two different sets of filters in the star simulator. In Fig. 10, the results using a bandwidth from 500 nm to 650 nm are plotted. When the star is focused close to the centre of the step, the visibility is 0.82±0.02, as shown in Fig. 10a. In Fig. 10b, the star is placed close to the edge and a second set of fringes begins to appear; the addition of the two visibility components is equal to 0.92. Fig. 10c shows the calibrated fringe measured when the star is focused approximately at the centre of the edge, so that the two components show similar contrast, and their addition is 0.91. From Fig. 10d, the addition of the two components is 0.90, and finally in Fig. 10e the fringe when the star is focused in the next step is plotted, and the visibility is 0.78. In order to compare the experimental results with the analytical predictions, we calculate the edge factor using the experimental parameters and Eq. (16). The average from the three measurements taken when the star is focused on the edge is multiplied by the edge factor to obtain the visibility. This resulting visibility is compared with the reference visibility obtained by averaging the two visibilities taken when the stars are focused on the steps. The obtained edge factor is 1.13±0.01. The average of the addition of the two contributions to the visibility is 0.91±0.02. After correction with the *f*-factor from Eqs. (15) and (16) a visibility of 0.80±0.02 is recovered. Given that this visibility is the same as the average visibility measured when the star is focused in the steps, we conclude that the visibility when the star is focused on the edge is completely recovered.

The results obtained using a bandwidth from 435 nm to 650 nm are plotted in Fig. 11. The average of the three measurements taken when the star is focused on the edge (Figs. 11b, 11c, and 11d) gives a total visibility of 0.81±0.02. From the value of the edge factor of 1.03 and, from

Eq. (15), a visibility of 0.79±0.03 is calculated. By averaging the two measurements of the fringe pattern when the star is focused on two steps (Figs. 11a and 11e) a visibility of 0.78±0.02 is obtained. This shows that in this case the visibility is again recovered.

**3.3.2 Experimental results with a 34 mm baseline**
For the experiment described in this section we changed the baseline of the interferometer to 34 mm. In order to use the same mirror, we also have to change the effective focal length of the set-up to 7.438 m to keep the relation between the depth and the width of the steps, as shown in Eq. (18). When changing the focal length, the positions of the stars in the focal plane change, and this allows to have a reference star on a step and another star on the edge between two steps. The results are depicted in Fig. 12. The visibility of the reference star, plotted in Fig. 12a, is 0.70±0.02. It is lower than previously due to dispersion effects. The calibrated fringe from the star focused on the edge between two steps is shown in Fig. 12b. The addition of the two visibility components results in a visibility of 0.80±0.02. The value of the edge factor remains 1.13±0.01 as in the previous case (Section 3.3.1). With this factor, a visibility of 0.71±0.02 is recovered, and the relative angular separation between the stars is $(3.0\pm0.1) \times 10^{-4}$ rad.

**4 Conclusions and discussions**
We have demonstrated a new approach to the field-of-view problem in optical interferometry. The main advantages of our approach are that a wide field of view is achieved in one shot and that it is a flexible method, compatible with image or pupil plane combination interferometers. We have shown experimental results that prove the feasibility of the concept. Fringes from an off-axis star, separated 1.07 arcmin from the on-axis reference star, have been obtained simultaneously and with a contrast similar to the on-axis one. The relative angle of the off-axis star has been retrieved and nicely corresponds to the expected value. In the case when a star is focused on the edge of a step, the modulated fringe pattern is split into two fringe sets, but the addition of the individual contrasts of the two sets, apart from a proportionality constant, yields the expected visibility as if the star was focused only on one step. We call this constant 'edge factor' and it can be directly obtained from the spectrum of the detected light and the mirror configuration. The theoretical analysis is in good agreement with the experimental results. The diffraction effects due to the edge have been studied using Fourier analysis to determine the limits of this approach; we have shown that it is valid when the diameter of the re-collimating lens is larger than 1.5-1.7 times the output pupil. The discontinuous nature of our wide-field solution does not imply that the acquired field of view is discontinuous. Using the simple algorithm described in this article, we show that a *continuous* wide field of view can be acquired in one shot with a Michelson pupil-plane interferometer.

Finally, we would like to point out that the star sample that we used in our experiment consists of only four well separated stars. One important remaining question is how the actual imaging process will be done once a more general wide field object is to be reconstructed. This requires a more extended reconstruction method based on joint deconvolution with space variant convolution kernel. Using an analytical method to calculate the wide field response with the addition of the staircase mirror in intermediate focus to equalise the field dependent *OPD* and by treating all fringe patterns jointly, we were able to simulate the reconstruction of a complicate wide-field linear structure on the sky.[16]

**Acknowledgements**

The authors would like to thank the Knowledge Center for Aperture Synthesis, a collaboration of TNO and Delft University of Technology, for financial support. We gratefully acknowledge fruitful discussions with Andreas Quirrenbach, for suggesting the analysis from Appendix A, and with Cas van der Avoort. We also acknowledge the technical support of Aad van der Lingen during the assembly and alignment of the set-up.

**Appendix A: Fourier analysis of the diffraction at the staircase step on the visibility**

The system we use for our theoretical analysis is represented in Fig. 3 and consists of a two-arm (*A* and *A´*) Michelson interferometer. At the entrance of each arm there is an aperture with a lens. In the focal plane of the lens in arm *A´* we place the staircase device formed by steps of depth *d*. When the focal image is exactly centred on the edge of the step a wavelength-dependent phase shift of $\delta(\lambda)$ is imparted to half the Fourier spectrum of the corresponding Point Spread Function (*PSF*). We collimate the beam from each arm again and then combine both beams. In order to study the effects that the phase shift $\delta(\lambda)$ has on the final detected power, we perform a 1-D simulation of the system.

We are simulating a Michelson interferometer and, after recombination, the apertures are both effectively centred at the origin and their diameters are equal and represented by *D*. The one-dimensional apertures are represented by the functions $A_1(x)$ and $A_1'(x)$,

$$A_1(x) = A_1'(x) = \mathrm{rect}(x/D), \tag{A1}$$

with:

$$\mathrm{rect}\left[(x-x_0)/d\right] = \begin{cases} 1 & |(x-x_0)/d| \leq 1/2 \\ 0 & |(x-x_0)/d| > 1/2 \end{cases}. \tag{A2}$$

The *PSF* of the system is

$$A_2(\lambda, f_x) = A'_2(\lambda, f_x) = \frac{D}{\lambda F} \operatorname{sinc}\left(\frac{D}{2} f_x\right), \qquad (A3)$$

with

$$f_x = \frac{2\pi x}{\lambda F} \text{ and } \operatorname{sinc}(x) = \frac{\sin(x)}{x}. \qquad (A4)$$

In the focal plane of arm $A´$ there is a staircase mirror that introduces a partial phase shift to half the spectrum of the *PSF*. It's represented by the function $S(f_x)$

$$S(f_x) = \begin{cases} 1 & f_x \leq 0 \\ \exp[-j\delta(\lambda)] & f_x > 0 \end{cases}. \qquad (A5)$$

The complex amplitude $A_3(x)$ in the far-field of arm $A$ is simply given by $A_1(x)$ and in arm $A´$ we obtain

$$A'_3(\lambda, x) = \frac{1}{2\pi} \int_{-\infty}^{0} A'_2(\lambda, f_x) \exp(jxf_x) df_x + \frac{1}{2\pi} \int_{0}^{-\infty} A'_2(\lambda, f_x) \exp[-j\delta(\lambda)] \exp(jxf_x) df_x. \qquad (A6)$$

Using Eq. (A3) we get

$$A'_3(\lambda, x) = A'_1(x) + \int_0^\infty \frac{1}{2\pi} \frac{D}{\lambda F} \operatorname{sinc}\left(\frac{D}{2} f_x\right) \{\exp[-j\delta(\lambda)] - 1\} \exp(jf_x x) dx. \qquad (A7)$$

Introducing Eq. (A7) in the computing code Maple we obtain the following expression:

$$A'_3(\lambda, x) = A'_1(x) + \frac{\{\exp[-j\delta(\lambda)] - 1\}}{2\pi} \left[\pi + 2j \operatorname{arctanh}\left(\frac{2x}{D}\right)\right], \qquad (A8)$$

and for the wavelength-integrated flux we obtain

$$I_3(x) = \int_{\lambda_0-\Delta\lambda}^{\lambda_0+\Delta\lambda} |A_3(x)|^2 \, d\lambda, \tag{A9}$$

$$I'_3(x) = \int_{\lambda_0-\Delta\lambda}^{\lambda_0+\Delta\lambda} |A'_3(\lambda, x)|^2 \, d\lambda, \tag{A10}$$

where $\lambda_0$ is the mean wavelength and $\Delta\lambda$ is the bandwidth.

Now we combine the beams from arm $A$ and $A'$. The diameter of the collimating lenses and the diameter of the beam combination lens are both equal and represented by $D_c$. In the combined focal plane the contribution of arm $A$, is given by ($D_c > D$)

$$A_4(\lambda, f_x) = A_2(\lambda, f_x), \tag{A11}$$

and for $D_c < D$ the result is

$$A_4(\lambda, f_x) = \frac{D_C}{\lambda F} \operatorname{sinc}\left(\frac{D_C}{2} f_x\right), \tag{A12}$$

The contribution from arm $A'$ is given by:

$$A'_4(\lambda, f_x) = A_4(\lambda, f_x) + \{\exp[-j\delta(\lambda)] - 1\} G(\lambda, f_x), \tag{A13}$$

where:

$$G(\lambda, f_x) = \int_{-\frac{D_C}{2}}^{\frac{D_C}{2}} \left[\frac{\pi + 2j \operatorname{arctanh}\left(\frac{2x}{D}\right)}{2\pi}\right] \exp(-jf_x x) \, dx. \tag{A14}$$

Finally the combined *PSF* is:

$$D(\lambda, f_x) = A_4(\lambda, f_x) + A'_4(\lambda, f_x). \tag{A15}$$

In the previous simulation we did not consider the effect of the finite apertures of the collimating lenses, but if we consider a finite diameter $D_c$, the high frequencies in function $A'_3(x)$ will be cutoff as an effect of the convolution of the Fourier transforms of the function itself and the function that defines the beam combination lens. As we are not integrating to infinity any more, the high frequencies are reduced, and this will affect the final measured visibility. For a collimating lens of 30 mm aperture, the same as in the experiment, the diffraction is not substantially affecting the measurements and we get back a visibility equal to 0.99. The results when the edge is not in the centre are similar.

Of course, in a real system, the collimating lens or mirror will have a size very similar to that of the output pupil of the telescopes. We have performed simulations for different ratios $D_c/D$. We overlapped the functions in the pupil plane, and modulating the optical path we observe two sets of fringes. We calculate the contrast from each fringe set, $V_1$ and $V_2$, and the power of each beam, $I_1$ and $I_2$, to calculate the modulus of the visibility of each fringe set, $\mu_1$ and $\mu_2$. Because the edge is in the middle of the *PSF*, in this case $V_1=V_2$ and $\mu_1=\mu_2$. Finally, we add both contributions and divide over the edge factor $f$, that was 1.13 in this case. Doing this we expect to recover a value approximately equal to one, or with an error of maximum 2%. The maximum power of the combined light in the absence of an edge has been normalised to unity. The results are presented in Table A.1.

The results show that by cutting the tails of the function, we are actually removing the coherent background that was producing the fluctuations calculated in the edge factor. Therefore, according to our calculations, when $D_c/D$ is smaller than approximately 1.5-1.7, one can not apply the procedure to calculate the expected visibility explained in this paper, because the error in the obtained result is larger than 2%. On the other hand, reducing the coherent background produces an increment in the visibility of each fringe set. This is due to the effect that the minimum power decreases slower than the maximum power, as plotted in Fig. 13. The position of the peaks of each fringe set relative to the reference remains unchanged.

Table A.1: calibrated modulus of the complex visibility calculated for different ratios of the collimating lens diameter over the output beam diameter.

| $D_c/D$ | 0.3 | 0.5 | 0.8 | 1 | 1.3 | 1.5 | 1.7 | 3 | 5 |
|---|---|---|---|---|---|---|---|---|---|
| $I_{max}$ | 0.20 | 0.34 | 0.55 | 0.71 | 0.74 | 0.75 | 0.755 | 0.77 | 0.77 |
| $I_{min}$ | 0.04 | 0.07 | 0.11 | 0.16 | 0.20 | 0.20 | 0.21 | 0.22 | 0.23 |
| $I_1$ | 0.07 | 0.12 | 0.20 | 0.25 | 0.25 | 0.25 | 0.25 | 0.25 | 0.25 |
| $I_2$ | 0.04 | 0.07 | 0.12 | 0.18 | 0.21 | 0.22 | 0.22 | 0.24 | 0.24 |
| $V_1$ | 0.68 | 0.67 | 0.66 | 0.62 | 0.58 | 0.57 | 0.57 | 0.56 | 0.56 |
| $\mu_1$ | 0.70 | 0.70 | 0.68 | 0.63 | 0.585 | 0.58 | 0.57 | 0.56 | 0.56 |
| $2\mu_1/f$ | 1.25 | 1.24 | 1.20 | 1.12 | 1.04 | 1.02 | 1.01 | 0.99 | 0.99 |

Figure 1: General scheme of Fizeau and Michelson interferometers: (a) the light is combined by refocusing the beams in a common focal plane giving as result spatial fringes. If $B_0=B/M$, being M the magnification of the system, the interferometer works as an homothetic mapper, and is in general called a Fizeau type. (b) The beams are combined with a beam splitter resulting in temporal fringes. This type is usually called a Michelson interferometer.

Figure 2: OPD equalization; the light from different directions in the field is focused on different steps that introduces an extra OPD as a function of the field angle.

Figure 3: Schematic representation of the interferometer. $A_1$ and $A_1$' are the diffracting apertures with lenses that produce images in the focal plane ($A_2$ and $A_2$'). A phase shift is introduced at $A_2$' to half the area of the focused beam. The two beams are then recombined using apertures $A_3$ and $A_3$' and the power modulation is measured at $D$.

Figure 4: Schematic representation of a source imaged on the edge between two steps in the staircase mirror.

Figure 5: Effect of the staircase mirror on an off-axis star that is imaged on the edge of a step: (a) position of the off-axis star, (b) the power, calculated by integrating the flux of the spot is plotted as a function of time. (c) The contrast from each set of fringes plotted in black squares and triangles, respectively. The black dots represent the addition of the contrast of the two sets of fringes corresponding to every baseline divided by the edge factor.

Figure 6: Analytical edge factor as a function of the baseline, for a spectral range of 500-650 nm.

Figure 7: Scheme of the experimental setup consisting of three blocks: the star simulator with the white light source and the star mask, the interferometer, with the staircase mirror in one of the arms, and the beam combiner with the delay line. A detector (PD) was placed in one of the outputs of the beam splitter and a CCD camera at the other to control the pupil position. The delay line (DL) was connected to a piezo driver and to a signal generator to modulate the path producing temporal fringes.

Figure 8: White light fringes when the mirror has the steps parallel to the baseline: (a) the on-axis calibrated fringe, (b) the off-axis calibrated fringe. The off-axis star fringe is out of the interferometric field of view; its peak is not observable. Detector power is plotted vs. sample number, that is equivalent to time.

Figure 9: White light fringes when the steps of the mirror are perpendicular to the baseline: (a) and (c) calibrated fringes from the on-axis stars, (b) and (d) calibrated fringes from the off-axis stars. The interferometric field of view has been extended and the peaks of the fringes of the off-axis stars are observed.

Figure 10: Experimental results of the effect of the edge on the visibility, with a bandwidth of 150 nm and a central wavelength of 575 nm; going from (a) to (e), the star is moved from the center of the central step to the center of the next step. In (a) and (e), because the star is focused in the center of the step, one fringe set is observed. In (b), as the star is moved close to the edge, a second fringe set appears, with contrast smaller than the other component. In (c), the star is focused approximately in the middle of the edge, both fringe sets present similar contrast. In (d), as the star is moved towards the center of the next step, the contrast of the first fringe set decreases and the contrast of the second one increases.

Figure 11: Same as in Figure 7, but for a bandwidth from 435 nm to 650 nm.

Figure 12: Experimental results of the effect of the edge on the visibility, with a bandwidth of 150 nm, a central wavelength of 575 nm, and a baseline of 34 mm: (a) the calibrated fringe from the reference star, (b) the calibrated fringe from the star focused on the edge between two steps.

Figure 13: Maximum power (black dots) and minimum power (black triangles) of the combined light, plotted as a function of $D_c/D$. When cutting the pupil, both the maximum and the minimum tend to zero, but the minimum decreases slower than the maximum, producing an increase in the contrast.

Figure 1

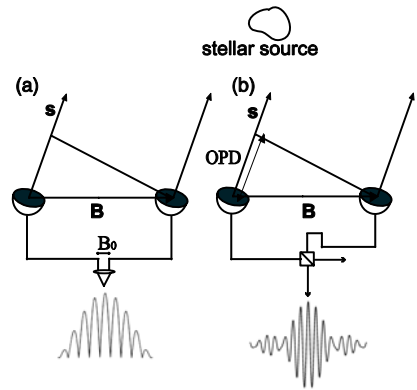

Figure 2

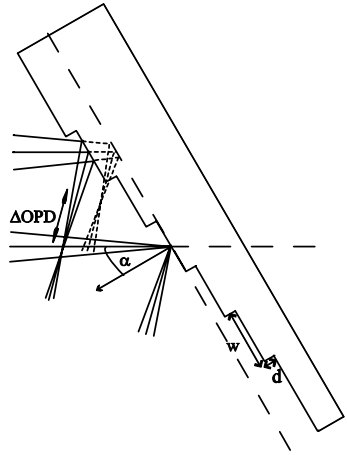

Figure 3

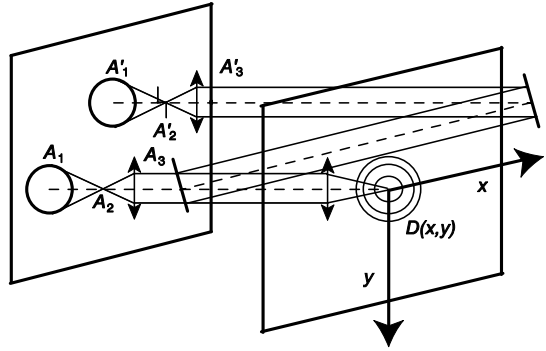

Figure 4

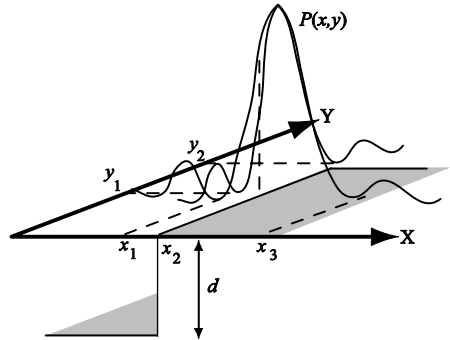

Figure 5

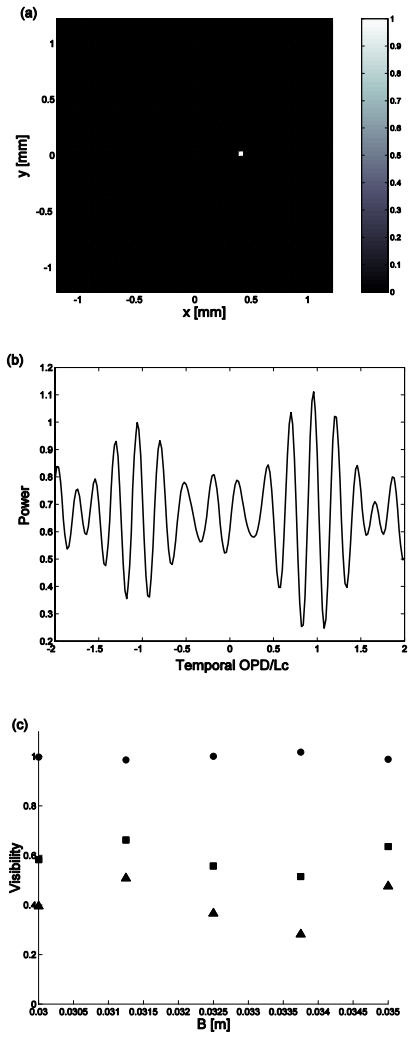

Figure 6

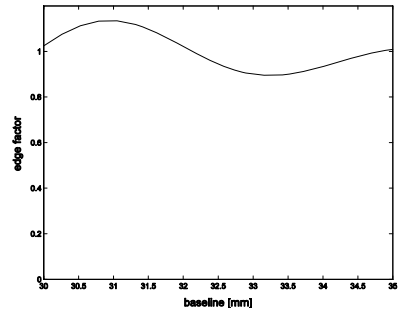

Figure 7

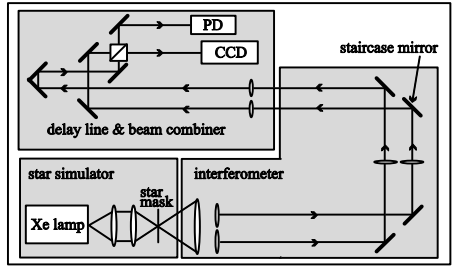

Figure 8

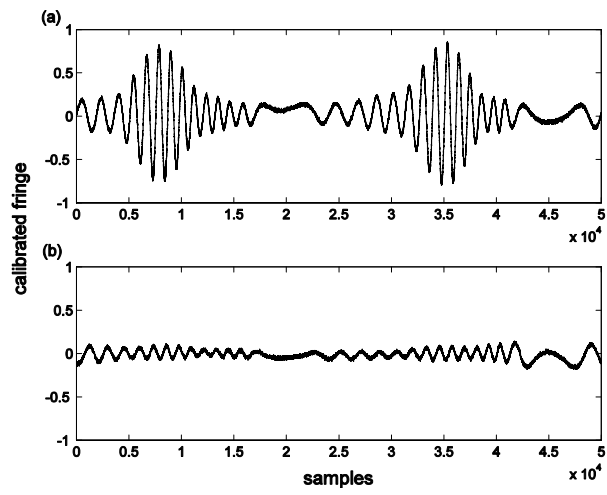

Figure 9

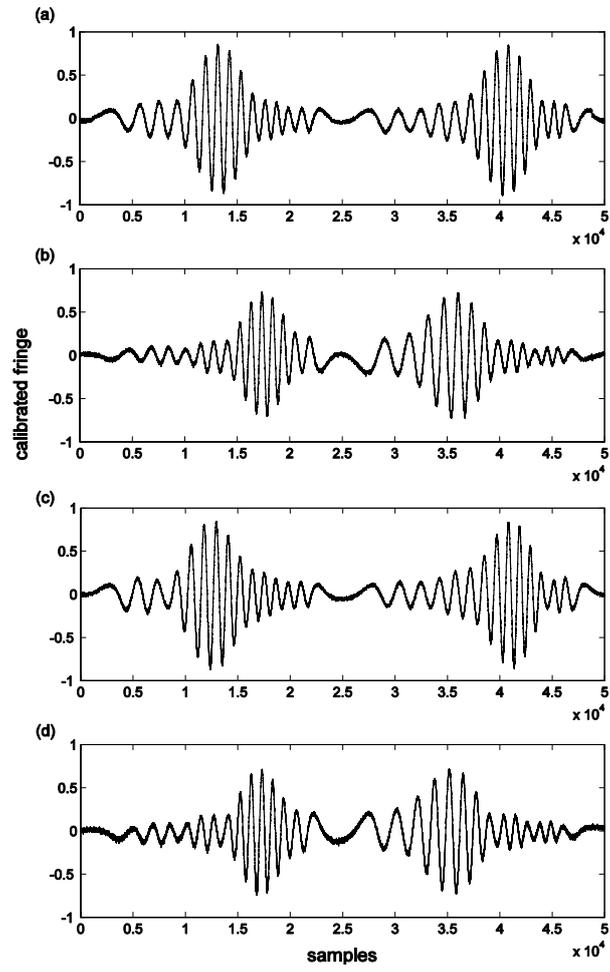

Figure 10

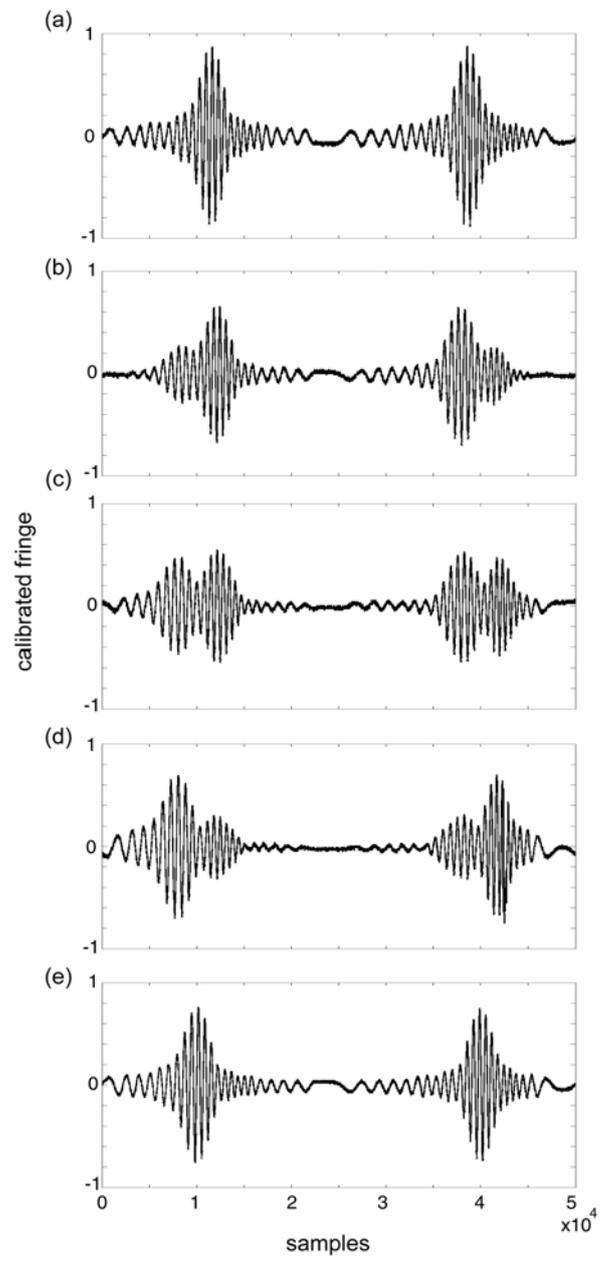

Figure 11

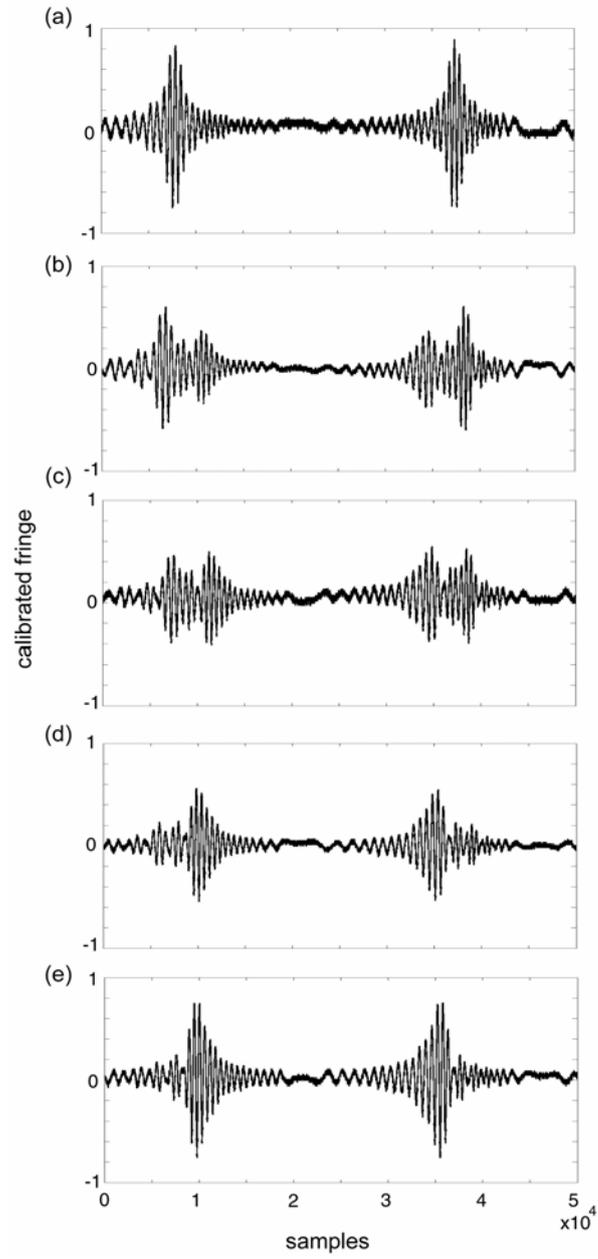

Figure 12

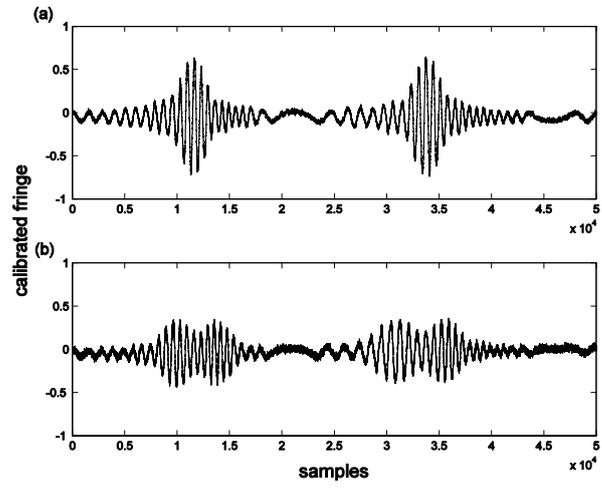

Figure 13

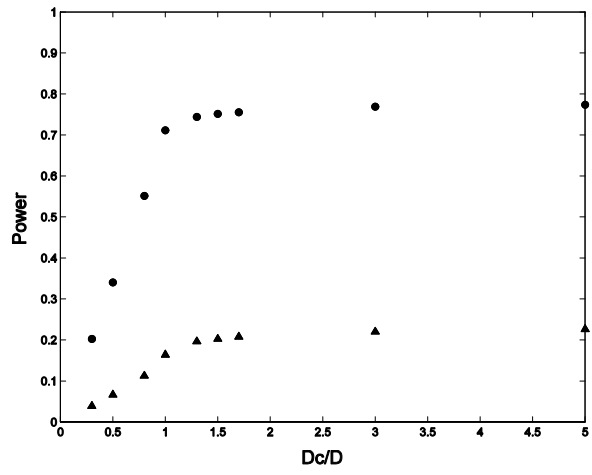